**The challenges of massification in higher education in Africa**

Authors :

Dr Kossi TEPE, University of Lomé (Lomé, Togo)

kossi.tepe@yahoo.fr

Dr Yann VERCHIER, Université of Technology of Troyes (Troyes, France), Associate Professor at University of Sherbrooke (Canada)

yann.verchier@utt.fr

Prof. Kokou YETONGNON, University of Bourgogne (Dijon, France)

kokou.yetongnon@gmail.com

**Abstract**

Like many developing countries, Togo faces the challenge of massification in higher education resulting from a large increase in the number of students enrolled in its public universities. Encouraged by the public authorities, with the support of the United Nations and Unesco, the number of students to be trained continues to grow to provide the country with qualified professionals and meet its socio-economic needs. The number of students in large groups (over 3,000 in some courses) raises issues of training quality and equity (availability of resources, reproducibility of content, study conditions, access to digital solutions, etc.). Access to this type of training requires special training conditions and infrastructures that are not always available in developing countries.

This article presents a qualitative study carried out with undergraduate students and teachers at the University of Lomé concerning teaching and learning conditions in large groups and a critical analysis of the solutions implemented by the university. This work can be transposed to other African countries with similar needs and will open the way to a solution analogous to intelligent classrooms for face-to-face courses.

**1- Introduction**

Demand for quality university education is growing rapidly in developing countries. Following the implementation of the United Nations' Sustainable Development Goals (SDGs) and the involvement of UNESCO (Wikipedia), (Tepe K., Yetongnon K.,2018), efforts are being made in almost all countries to provide access to quality university education for populations. While secondary education is no longer a major challenge today in most countries, university-level training poses many challenges for developing countries concerned with training senior executives capable of tackling sustainable development problems and meeting the requirements of financial institutions such as the African Development Bank (ADB), the International Monetary Fund (IMF) and others (Pierre-Jean Loiret, 2008).

To meet this demand, almost all African countries have created public universities since the 1970s (lecames.org, 2018); but their numbers remain insufficient. This inadequacy has encouraged the emergence of private higher education institutions in several countries. However, the costs of these private institutions are very high in relation to the socio-economic profiles of students. In addition, they suffer from many criticisms linked to unsuitable teaching content and teaching teams' competence level.

The above factors explain the ever-increasing demand for training at African public universities. Some university courses are more in demand than others, creating large groups (3,000 students or more) in teaching some units. As a result, public universities are struggling to manage this high demand with the limited resources at their disposal.

Hosting large groups of undergraduates at public universities requires special conditions and adaptations (in terms of space and time) to ensure high-quality supervision of these students. The resources required by the teaching teams must not only be substantial but must also include additional technical resources to cope with this high demand for university training. This is the only way to ensure that training remains of the highest quality, meets the country's needs, and is recognized internationally.

Managing large groups is becoming a major challenge to which African universities must find solutions. The solutions adopted by several universities include building large classrooms for face-to-face courses and offering courses online for distance learning. These solutions are inadequate, however, as the capacity of classrooms cannot be increased indefinitely without posing new problems for audience management. On the other hand, distance learning poses new problems from the point of view of both students and teachers. These problems must be carefully studied to find appropriate solutions (access to resources, motivation, student monitoring, and certification of prior learning) (Jézégou, A., 1998).

Given the high demand for training, it is necessary to provide African public universities with adequate infrastructure for hosting and managing large groups. In this article, we analyze students' perceptions of large-group learning activities and the hybrid model of large-group management, which combines the advantages of both face-to-face and distance learning.

**2. State of the art**

Information and communication technologies have fostered the emergence and development of pedagogical innovations that affect methods of disseminating or acquiring knowledge, the types and size of target audiences, and the costs of education. Several large-scale international and national projects have been proposed to improve the capacity of African universities to meet the growing demand for higher education. These projects, generally based on ICT tools, aim to overcome the lack of infrastructure, teaching resources, and qualified teachers in certain fields. Most of these projects have not produced the expected results, and new pedagogical methodologies for mass training are being sought all over the world (Jézégou, A.,1998); (Savoie-Zajc, L., & Karsenti, T., 2018); (Mitchell A., .2014); (Arbaugh, J. B., & Rau, B., 2007). There are a growing number of mass training models centered on information and communication technologies. They include:

- **MOOCs (Massive Open Online Courses)** and e-learning are distance learning courses aimed at the general public and offering resources for acquiring new knowledge and developing personal or professional skills, (Haipinge and Kadhila, 2021).

- **E-learning** enables the delivery of training courses to a target audience who can follow the course individually or collectively (e.g., planning and scheduling courses, lab and discussion sessions with online presence of teachers). Several prestigious universities have adopted these new educational technologies to offer online, fee-based training courses equivalent to the degree courses offered by these institutions in initial training (Hone, K.S.et al., 2016); (Wu, B. and Chen, X. 2017); (Margaryan, A., et al., 2015). As part of the United Nations' ODD4 initiative, the International Institute of Online Education uses new technologies to provide learning platforms for higher education teachers. IIOE (www.IIOE.org) is a global initiative created by UNESCO-ICHEI in collaboration with higher education partner institutions in China and other developing countries in Asia-Pacific and Africa, as well as corporate partners. It aims to build the capacity of the partner institutions and expand access to quality higher education. The target audience for MOOCs and e-learning may include professionals or self-taught individuals who can tailor courses to suit their availability. Examples of popular MOOCs include Coursera (www.coursera.org), Stanford (www.Stanford.edu), and MIT (www.mit.edu). Some courses lead to professional certificates, such as those issued by ORACLE (www.oracle.com), CISCO (www.cisco.com), Microsoft(www.microsoft.com), IIOE, etc.

- **Virtual universities** allow large groups to access high-quality university training at low cost. At the international level, several large-scale projects were proposed in the 90s, notably the African Virtual University (AVU) project, initiated by the World Bank and the ADB (Pierre-Jean Loiret, 2008). The goal is to transfer technology from universities in northern countries (e.g. University of Galway, Virginia Polytechnic University, University of Toulouse, University of Brussels) to french-speaking (Benin, Burundi, Mauritania, Niger, Senegal) and english-speaking (Ethiopia, Ghana, Kenya, Uganda, Tanzania, Zimbabwe) African higher education establishments. The pedagogical model is based on creating learning centers in African institutions, from which learners can follow a selection of courses given in the partner institutions via the Internet. The AVU's business model is based on reduced public funding and greater financial participation by students. The project has not been as successful as expected, and none of the participating universities have adopted its video broadcasting principle (www.campus-teranga.com). In a similar vein, the CVA project (African Virtual Campus piloted by UNESCO and its partners), launched in November 2008, aims to bring together 5 West African countries (Benin, Cape Verde, Côte d'Ivoire, Gambia and Senegal) to provide the sub-region with a common program through a network of online production and teaching centers in West Africa (with a precise model for copyright and intellectual property). This project also failed to deliver the expected results (Patrick Vigliano,2009). UoPeople (University of people) (www.uopeople.edu) is a private virtual university set up in 2009 in partnership with several American and Canadian universities, offering a limited number of bachelor's and master's degrees in business, health science, education, and other fields. The teaching methodology is based on online course delivery to groups of 20 to 30 students, assisted by an instructor. Registration fees are virtually free, and the resources required (books, videos, recordings, etc.) for the courses are provided via the OER (Open Educational Resources, www.oercommons.org) and unesco.org (www.unesco.org) sharing platforms.

At the national level, several projects have been proposed, including the "Universités Virtuelles du Sénégal (UVS) (www.uvs.sn)", Zimbabwe ZOU (Zimbabwe Open University, www.zou.ac.zw), Pakistan (Virtual University of Pakistan, www.vu.edu.pk), and India IGNOU (Indira Gandhi National Open University, www.study badshah.com). The UVS project, created in 2013, offers an innovative approach by considering characteristics specific to developing countries, namely inadequate infrastructure, economic, financial, sociological, and ultimately political factors (Bichat, 2012). The teaching methodology is based on face-to-face and distance learning courses. The strategy adopted is to offer new learners general face-to-face courses and then move to online teaching once students have acquired the autonomy, maturity, and digital skills to follow individual or group modules online. A key component of the teaching methodology is the ENO (Espace Numérique Ouvert, www.uvs.sn/eno-de-luvs/presentation-eno), which gives students access to teaching resources (course videos, digital tools, etc.).

- **Videoconference-based training** gained in popularity during the COVID 19 health crisis, and enabled institutions around the world to maintain an acceptable level of training programming and delivery. Video Conferencing tools (ZOOM, TEAMS, MEET, WEBEX etc.) have made it possible to move course programming into a virtual space that learners can access collectively from different locations. These solutions have met with some success but require further study and analysis to determine whether the level of teaching and assessment is comparable to that of traditional training.
- **Training courses based on video courses on the Internet** (YouTube platform), enabling tutorial videos to be put online and shared without the objective of a global training course leading to a diploma (www.youtube.com). These courses are intended more for self-training on specific subjects, with no certification of content or validation of prior learning. Content producers tend to aim for a large number of subscriptions as a reward. It's a free, autonomous acquisition of knowledge to complete an education or make up for shortcomings in given subjects, without any training costs. The quality of the trainers and the accuracy of the content are not guaranteed. The target audience is general, global, and made up of any Internet user looking for educational videos on various academic or non-academic subjects.

**3- Study scope**

TOGO has two main public universities, the University of Lomé and the University of Kara. For several years now, these two universities have been experiencing an ever-increasing demand for access to university training (post-baccalaureate) by first-time learners from secondary schools. This demand is due to the need to achieve sustainable development in line with the United Nations SDGs (SDG4), (Tepe K., Yetongnon K., 2018), (Pierre-Jean Loiret, 2008) and an unavoidable imperative to train senior executives in the best conditions for quality teaching. This situation is a constant concern for university managers who must deal with numerous organization and quality problems in managing large groups. The example of the University of Lomé (UL) will be used to illustrate the motivation for our study and proposed solution.

The Table 1 below shows a ten-year trend in first-year enrolments in the faculties of literature, economics and management, law, and science at the University of Lomé. The table highlights the uneven distribution of large numbers of students across the different disciplines, necessitating a search for appropriate solutions for managing large numbers of students, particularly for lectures, discussion groups and lab sessions.

| Type of faculty | **Year 2011-2012** | Year 2012-2013 | Year 2019-2020 | Year 2020-2021 | **Year 2021-2022** |
|---|---|---|---|---|---|
| Literature, Languages and Arts | 2 919 | 3 153 | 5 085 | 4932 | 5 484 |
| Faculty of Economics and Management | 2 620 | 3 273 | 5 415 | 7080 | 6 270 |
| Law | **3 876** | 2 246 | 2 772 | 2828 | **2 921** |
| Faculty of Science | **2 290** | 1 877 | 3 083 | 4019 | **3 347** |

**Table 1**: Evolution of semester 1 large group enrolments at UL over ten academic years with some reference years (academic data of University of Lomé; 2021)

The authorities at the University of Lomé have implemented several measures to facilitate teaching the large groups. They are based on increasing the material infrastructures and human resources, mainly:

- Constructing several large 1000 and 1500-seat amphitheaters

- Setting up a wireless campus network to bring together the universities of Lomé and Kara, and hospital centers to share resources. Students at the different sites can access the Internet via a single account.

- Organization courses in parallel sessions for teaching units with very high demand

- Hiring more teachers for the repetition of courses (parallel sessions). This requires harmonizing the content followed by different audiences for the same curricula.

- Setting up work and rest areas on campus to facilitate mobile work with access to the campus network.

- Putting courses online enabled not only the continuity of students' personal work, but also the management of distance courses by teachers. The creation of the online course platform and the training of the various users (students, teachers and technical staff) were stepped up, especially in the context of the COVID-19 pandemic.

- Scheduling of hybrid (distance and face-to-face) or 100% distance teaching during the pandemic.

The initial feedback from stakeholders on the measures on training at the University of Lomé (students, lecturers, technical and administrative staff, etc.) shows that the results are not satisfactory and that more needs to be done. In some teaching units, the current 1000 and 1500-seat lecture rooms do not have enough seats for all the students, who constantly compete for the available places. This situation has led to unrest and anarchic occupation of the premises. Some students have to reserve or occupy available seats in the lecture hall hours before classes start. These disruptions are more pronounced when two classes are scheduled in the same room at short intervals. Moreover, in some large lecture theaters, students have difficulty following lectures due to technical faults or poor use of teaching aids (microphones, sound systems and video projection).

**4- Study methodology**

To find solutions for managing large groups of students, a study was carried out to gather the opinions of the various stakeholders at the University of Lomé. The aim was to analyze the interest and effectiveness of the current solutions, as well as the possibility of considering other, more suitable solutions using ICT tools. The study adopted a mixed-method approach, combining open interviews and questionnaires designed to obtain quantitative results. It consists of :

- Questionnaires with open and closed questions for students and teachers.

- Data collection through open interviews. These exchanges aim to collect additional information that teachers and students would not have provided in the survey forms. It should be noted that at the University of Lomé, the status of lecturers does not always allow for free expression, thus, some of the results obtained from the questionnaire must be supplemented by anonymous interviews.

The teacher surveys aimed to find out how they felt about teaching large groups (interaction, fair treatment, working comfort, etc.) and to understand how they had invested in the various distance learning solutions implemented during COVID period. It should be noted that, as the profile of the teachers varied widely (tenured, part-time, industrialists, etc.), the questionnaire began with a set of questions designed to define the profile of each respondent.

Student surveys focused mainly on the transition from high school to university regarding working conditions (group size, relationship with the teacher, personalized student support, etc.). An important part of the questionnaire focused on their connection tools (network quality, computer equipment for taking courses, etc.) and their ability to have a suitable workspace when taking distance learning courses.

The questionnaires were sent to 1,800 students in the bachelor's program, the majority of whom were in large groups, and to 106 teachers who made lectures in various large groups. Eighty-six responses were received from teachers and 1,702 from students. The results are presented and interpreted in the following section. Seventy open interviews were conducted with teachers from various teaching teams involved in the major bachelor's and master's degree groups.

**5- Results and interpretation**

**5.1. Teachers survey**

The 86 teachers who took part in the survey agreed that an effective solution needs to be found for managing large groups of students in public universities. 58% of the survey respondents suggested dividing large groups into smaller ones for lectures. In comparison, 16.7% suggested finding a solution to relocate online courses entirely. The same percentage thought that a solution that offered the same courses online and face-to-face, to enable those who couldn't access lecture halls to take them online, would be more suitable. Informal discussions with lecturers proposing a solution for making courses available online raise the issue of technological solutions for protecting access to copies of courses and authorizing only students enrolled in these courses to listen to and view the sessions. They also raise problems of mastery of the new ICT tools that enable distance courses to be run, as noted elsewhere by the studies of (Erstad and al.,2021) and which has even led to the proposal of a basic technological skills test tool in teacher recruitment in Belgium by the work of (Tondeur et al.,2017).

50% of those surveyed thought that a reasonable size for a large group would be 50 students, 16.7% thought that a large group would contain 100 students, the same percentage thought that a large group could contain up to 300 students and only 8.3 thought that a large group could even contain up to 500 students.

With maximum sound comfort, 16.7% of opinions are divided in favor of the best possible supervision of students of a size of 100, 200, and 300. 50% believe that 50 students should not be exceeded. Further informal discussions with the majority of teachers led to the conclusion that, for practical purposes and skills assessment, the size of student groups that can be easily managed should not exceed 50 learners for reasons of material organization.

50% of the survey estimates that in one day, for efficiency in teaching the same course for 2 hours, only 2 groups will be appropriate, 41.7% estimates that up to 3 groups of students can be taught in one day. No survey expression for 4 groups in one day. During informal discussions, a majority of teachers who had already repeated the same lessons in different sessions felt that the content of the lessons was never identical. Indeed, students' questions could lead to other developments in some sessions, and sometimes fatigue or good humor can also lead to not faithfully giving the same lessons in different groups. This could have a negative impact on the acquisition of skills and the assessments of the groups put together.

83.3% of the survey affirmed that the distance learning arrangements implemented during the coronavirus pandemic were ineffective. Informal discussions with teachers suggest unpreparedness for distance learning courses. Some face-to-face practices were used online, for example teachers sought to control the presence of remotely-connected students so they could dictate lectures or interrogate them using virtual classroom ICT tools such as Discord and others. They sometimes uploaded materials to Moodle for access and appointments for live explanations. Teachers also had difficulty adapting their practices to distance-learning ICT tools, as in similar cases elsewhere. (Blume, 2020) and (Greenhow et al., 2021).

To effectively achieve the pedagogical objectives of large group teaching, 50% believe that face-to-face training is needed, and 50% believe that face-to-face training should be combined with distance learning (i.e. a hybrid system).

**5.2.** Student **survey**

1702 responses were collected. The selected responses provide the following information and analysis:

- The majority of students surveyed are aged between 21 and 30 (66.7%), with only 33.3% aged between 15 and 20. Most of the 88.9% surveyed are at Bachelor's level, the level for which students are in the majority, with the problems posed to public universities by the management of these large groups. 39% of respondents believe that university studies are more difficult than high school studies.
- 72.2% of respondents are in groups of over 100 students.
- 80.6% of respondents are in groups of 1,000 to 2,000 students per class for certain teaching units.
- 58.3% of those surveyed said that the problem with large groups was communication, and 16.7% felt that there were also difficulties in carrying out the learning activities proposed by the teacher.

- Half of those surveyed do not have a workspace at home to facilitate access to the distance learning course. Among those who do have workspaces, the majority feel that working conditions do not allow for the levels of concentration that could be achieved in the classroom. This was particularly noticeable during the palliative distance learning courses organized during the coronavirus pandemic, with disturbing noises coming from the students' workplaces.
- 69.4% of those surveyed did not own a computer before attending university. Most of the courses taken before university were not based on using ICT tools. Most respondents felt uncomfortable with distance learning without a teacher alongside them, as in initial training, which discourages learning and personal development in the course modules offered online. These are some of the additional responses obtained from learners in informal discussions. This is very real when we see the work done elsewhere in e-learning solutions (Hone, K.S., El Said, G.R. 2016); (Arbaugh, J. B., & Rau, B. ,2007); (Pierre-Jean Loiret, 2008).
- For distance learning solutions, 72.2% believe they can take courses from their smartphone. 27.8% of students surveyed thought they could not for reasons of user-friendliness and familiarity with handling smartphones. This additional information was obtained informally from students for experiments carried out during the coronavirus pandemic.

**6- Conclusion**

Overall, the study points to the need to find an effective solution to the problems of managing large groups in countries with the same problems as Togo, especially for undergraduate or bachelor's degree students. Solutions based exclusively on distance learning will not be appropriate, but a face-to-face solution, or one that can combine distance and face-to-face teaching, will be more suitable. We also need to find a solution that enables courses to be automatically recorded and made available as needed to students who have not been able to attend a training module in which they are legally enrolled, and or repeat teaching sessions. Indeed :

- for students, training in a virtual environment presents a greater risk of discouragement and dropping out. These courses require technological, organizational and time management skills that most new university students do not possess. Other difficulties are linked to the lack of financial means and infrastructure to access distance learning courses. These include problems of electricity, Internet access, lack of personal workspace at home, and the acquisition of computers or equivalent devices (smartphones, tablets, etc.).

- for teachers, the virtual distance learning environment does not allow effective monitoring and evaluation of teaching and learner participation. The possibility of managing several groups of students in a single day has limits for effective teaching, and the solution to be found must take this fact into account. In addition, with informal contacts, some teachers have the feeling of being dispossessed of their teaching for distance learning courses with the online availability of courses. The lack of mastery of new ICT tools represents a major obstacle for other teachers, who may find it difficult to adapt classroom teaching to distance learning models.

- The solutions to be envisaged must be similar to intelligent classroom solutions (Cisse H., 2022), where teachers and learners must use ICT tools and innovative technologies at lower cost to facilitate course accessibility. The implementation of lessons with the possibility of reproducing the lessons delivered or replaying the same content without further effort by teachers with the combination of audio and video supports as proposed in education by (Nagro & Cornelius, 2013 ), (Pouta et al., 2021) and (Sommerhoff et al., 2022) would be more appropriate.